\documentclass[twocolumn,aps]{revtex4}

\newcommand{\beq}{\begin{equation}}  
\newcommand{\eeq}{\end{equation}}
\newcommand{\bea}{\begin{eqnarray}}
\newcommand{\eea}{\end{eqnarray}}

\begin{document}
%\draft

\title{ The minimal supersymmetric grand unified theory}

\author{
Charanjit S. Aulakh$^{(1)}$, Borut Bajc$^{(2)}$, 
Alejandra Melfo$^{(3)}$, Goran Senjanovi\'c$^{(4)}$ 
and Francesco Vissani$^{(5)}$}
\affiliation{$^{(1)}$ {\it Dept. of Physics, Panjab University, 
Chandigarh, India}}
\affiliation{$^{(2)}$ {\it J. Stefan Institute, 1001 Ljubljana, Slovenia}}
\affiliation{$^{(3)}$ {\it Centro de Astrof\'{\i}sica Te\'orica, 
Universidad de Los Andes, M\'erida, Venezuela}}
\affiliation{$^{(4)}${\it International Center for Theoretical Physics, 
Trieste, Italy}}
\affiliation{$^{(5)}${\it INFN, Laboratori Nazionali 
del Gran Sasso, Theory Group, Italy}}

\begin{abstract}
We show that the minimal renormalizable supersymmetric SO(10) GUT with 
the usual three generations of spinors has a Higgs sector consisting 
only  of a ``light'' ${\bf 10}$-dimensional and ``heavy'' ${\bf 126}$, 
$\overline{\bf 126}$ and ${\bf 210}$ supermultiplets. The theory has 
only two sets of Yukawa couplings with fifteen real parameters 
and ten real parameters in the Higgs superpotential. It accounts 
correctly for all the fermion masses and mixings. The theory predicts 
at low energies the MSSM with exact R-parity. It is arguably the minimal 
consistent supersymmetric grand unified theory. 

\end{abstract}
\maketitle

{\it A. Introduction}\hspace{0.5cm} 
After more than twenty years of low energy supersymmetry and 
grand unification the minimal supersymmetric Grand Unified Theory (GUT) is 
still commonly assumed to be based on the
SU(5) gauge group with three generations of $\overline{5}$ and
$10$ dimensional matter supermultiplets, and  $5_H$, 
$\overline{5}_H$ and $24_H$ 
-dimensional Higgs supermultiplets \cite{Dimopoulos:1981zb}. 
The ``Higgs'' superpotential contains 4 complex parameters. 
It has two sets of Yukawa couplings, and as is well-known
predicts equal down quark and charged lepton masses at the GUT scale,
relations which only work for the third generation and fail badly for
the first two. This can be corrected e.g.\ by $1/M_{Pl}$ terms, but 
at least two new sets of Yukawa couplings should be included.

On top of that, this minimal GUT is plagued by R-parity violating
couplings. If these are set to zero, neutrinos 
are predicted to be massless, which in view of atmospheric
and solar neutrino data seems an untenable position.
To correct for that, there are the following options:

\noindent (i)~add the effective, non-renormalizable, couplings 
$f_{ij}\overline{5}_i\overline{5}_j5_H5_H/M$, where $M\gg M_{GUT}$. 
The natural choice $M=M_{Pl}$ fails, since the data 
strognly suggest $M<M_{GUT}$. 

\noindent (ii)~add the right-handed neutrinos, which means their 3 masses 
 and 9 complex Dirac Yukawa couplings.

\noindent (iii)~add $15_H$ and $\overline{15}_H$ dimensional 
Higgs superfields with 6 complex Yukawa couplings, 
$f_{ij}\overline{5}_i\overline{5}_j 15_H$, 
and 4 more complex couplings among $15_H$, $\overline{15}_H$, 
$5_H$, $\overline{5}_H$ and $24_H$. 

Let us count the number of parameters for any of these possible versions 
of the minimal realistic supersymmetric SU(5) theory. We can diagonalize 
down quark (charged leptons) Yukawa, which means 3 real parameters. 
The symmetric up-quark Yukawas with the freedom of a global U(1) 
rotation of $5_H$ give us $6\times 2-1=11$ real parameters. 
With the $\overline{5}_H$ and $24$ fields in $W_H$ we can 
redefine two phases, thus $4\times 2-2=6$ real parameters in 
$W_H$. So, with massless neutrinos we have already 
\begin{center}
{\em $14(W_Y)+6(W_H)+1(gauge)=21 $ real parameters.}
\end{center}
Add neutrino masses and you get

\noindent (i)~$6\times 2=12$ more real parameters in 
$W_Y\;\to\;33$ real parameters in total;

\noindent (ii)~$3+9\times 2=21$ more real parameters in 
$W_Y\;\to\;42$ real parameters in total;

\noindent (iii)~$6\times 2=12$ more real parameters in 
$W_Y$, $4\times 2-2=6$ real parameters in $W_H$ and then the 
total is $39$ real parameters.

Strictly speaking, (i) should be discarded; a realistic theory has at 
least $39$ parameters. We keep (i), though, in order to emphasize the 
minimality of the SO(10) theory which we discuss below.

It is instructive to compare SU(5) with the minimal supersymmetric 
standard model (MSSM) with massive neutrinos. By this we mean effective 
neutrino mass operators of the type (i) mentioned in the case of SU(5), 
which amounts to effective Majorana masses for neutrinos. As usual, 
we assume R-parity (otherwise you have many more parameters). 
We have 6 quark masses, 3 quark mixings and 1 CP 
phase, 6 lepton masses, 3 lepton mixings and 3 CP phases, so in total 
22 real parameters in $W_Y$. With 3 gauge couplings and 1 real parameter 
($\mu$ term) in $W_H$, we get in total
\begin{center}
{\em 26 real parameters in {\rm MSSM}.}
\end{center}
Notice that we did not count the soft terms, they ought to be included 
separately, but they are present in any theory.

In an analogous manner we construct in this letter 
the minimal renormalizable
supersymmetric SO(10) theory. We find, much to our surprise, that the
theory has 26 real parameters, equal to the MSSM, and much less than 
the minimal SU(5). And yet it consistently describes 
all the low-energy phenomena, in particular
the fermionic masses and mixing angles. It is also a theory of
R-parity and it predicts its conservation to all orders in
perturbation theory. 
In short, it is the minimal
supersymmetric grand unified theory based on a single gauge group
without additional ad-hoc symmetries. 
The further tests of the theory will be provided by $d=5$ proton decay, 
leptogenesis and flavour violation processes.

At the first glance, this 
result of SO(10) being a minimal theory, more economical, simpler than 
SU(5), may appear as a contradiction in terms. After all, SU(5) is a 
smaller gauge group. Here, it may be useful to recall a lesson from 
the SO(3) model of leptons of Georgi and Glashow \cite{Georgi:cj}. 
In spite of being a smaller gauge group than the SU(2)$\times$U(1) 
theory, it is a much more complicated theory with many more parameters 
in the Yukawa sector. In this sense, the SM is the {\it minimal} 
electro-weak theory, not just the minimal correct one.

In this work, we describe the salient features of 
the minimal SO(10) theory. 
We pay special attention to the symmetry 
breaking analysis and the study of fermionic
masses and mixings. 

{\it B. The theory}\hspace{0.5cm} 
We construct now a renormalizable SO(10) theory. The reader may ask why 
necessarily renormalizable. Why not take $1/M_{Pl}$ terms? After all they 
could play a nontrivial role in fermion masses and mixings. However in 
this case we loose both minimality and predictivity. We have a nice 
example of the principle of renormalizability: the minimal standard model 
itself. Both its great phenomenological success and its eventual failure 
in predicting massless neutrinos are very useful. At the same time you 
have a precise starting point and a window toward the physics at high 
energies. We believe thus that it is rather important to extend this to 
the theory of grand unification and test it to the bitter end. Let 
experiment decide its fate and hopefully open a window for new physics 
even beyond, be it strongly interacting new dynamics, quantum gravity, 
superstrings, whatever. 

It is well-known that this requires ${\bf 126}$ ($\Sigma$) and 
$\overline{{\bf 126}}$ ($\overline\Sigma$) supermultiplets:
${\bf\overline{126}}$ alone produce right-handed 
neutrino masses and does not lead to anomalies, 
but its vacuum expectation value
(VEV) leads to a non-vanishing D-term, and therefore 
to supersymmetry breaking; adding ${\bf 126}$, 
supersymmetry can be preserved at a high scale
by assuming $\langle \Sigma\rangle =\langle \overline\Sigma\rangle $. 
These VEVs are SU(5) singlets, so they
leave SU(5) unbroken: thus, more 
Higgs fields are needed. The minimal choice is the four-index antisymmetric 
representation ${\bf 210}$ ($\Phi$), 
that was first considered in \cite{mc}. 
Notice that neither ${\bf 45}$ 
nor ${\bf 54}$ by themselves can do the job at the renormalizable 
level, both are needed \cite{Aulakh:2000sn}, 
leading to more parameters in the superpotential. 
In short, the minimal theory consists of $\Phi,
\Sigma $ and $\overline\Sigma$. The superpotential involving those heavy
fields is
\bea
W_H &=& \frac{m}{4!}\Phi_{ijkl}\Phi_{ijkl} +
\frac{\lambda}{4!}\Phi_{ijkl}\Phi_{klmn}\Phi_{mnij} \nonumber \\
&+ &\frac{M}{5!}\Sigma_{ijklm}\overline\Sigma_{ijklm} +
\frac{\eta}{4!}\Phi_{ijkl}\Sigma_{ijmno}\overline\Sigma_{klmno}
\label{superpot}
\eea
and it has only 4 independent parameters. In other words,
although the representations are large, the number of parameters is
small; the theory is simple and economical. We argue in passing
against the usual phobia of large representations: it is not size that
matters!

It is convenient to decompose the Higgs superfields under the
Pati-Salam maximal subgroup SU(2)$_L\times$SU(2)$_R\times$SU(4)$_C$ 
(for a useful connection between SO(10) and Pati-Salam see 
\cite{Aulakh:2002zr}):
\begin{eqnarray}
\Phi \equiv {\bf 210} & = & (1,1,15) + (1,1,1) + (1,3,15) \nonumber \\
& + &(3,1,15) + (2,2,6) + (2,2,10) + (2,2,\overline{10}) \nonumber \\
\Sigma \equiv  {\bf 126} & = & (1,3,\overline{10}) + 
(3,1,10) + (1,1,6) + (2,2,15) \nonumber \\
\overline\Sigma \equiv \overline{{\bf 126}} & = & (1,3,10) + 
(3,1,\overline{10}) + (1,1,6) + (2,2,15) \nonumber 
\end{eqnarray}
The physically allowed VEVs we call $p$ for $(1,1,1)$, $a$ for 
$(1,1,15)$, $\omega$ for $(1,3,15)$ and $\sigma$ and 
$\overline\sigma$ for $(1,3,\overline{10})$ and $(1,3,10)$. 
After some straightforward computation, the superpotential as a 
function of these VEVs becomes 
\begin{eqnarray}
W_H&=&m\left(p^2+3a^2+6\omega^2\right)
+2 \lambda\left(a^3+3p\omega^2+6a\omega^2\right)\nonumber\\
&+&M\sigma\overline\sigma+\eta\sigma\overline\sigma\left(
p+3a+6 \omega\right)\;.
\end{eqnarray}
Vanishing of the D-terms ($\Rightarrow \sigma=\bar\sigma$) and the F-terms
determines the supersymmetric vacua. The SU(5) symmetric vacuum
$p=a=\omega$ and the L-R symmetric one $p=\omega=\sigma=0$,
$a=-m/\lambda$ have simple expressions, while the standard model vacuum is
\bea
p=-\frac{m}{\lambda}\cdot \frac{x (1-5 x^2)}{(1-x)^2},\ a=
-\frac{m}{\lambda} \cdot \frac{(1 -2 x-x^2)}{(1-x)}, \nonumber  \\ 
\omega=\frac{m}{\lambda} \cdot x,\
{\sigma^2}= \frac{2 m^2}{\eta\lambda} \cdot
\frac{x (1-3 x) (1+x^2)}{(1-x)^2},
\eea
where $x$ is a solution of the cubic equation:
\begin{equation}
3 -14 +15 x^2-8x^3 =
(x-1)^2 \ {\lambda M}/{\eta m}
\end{equation}
The most typical case has single step breaking. E.g., when
$\lambda M\sim \eta m$, we get $p\sim-0.27\; m/\lambda$, $a\sim-0.67\;
m/\lambda$, $\omega \sim -0.21\; m/\lambda$, $\sigma \sim 0.51
m/\sqrt{\eta \lambda}$, that corresponds to an approximate single step
breaking of SO(10)$\to$MSSM.  However, in some cases $\sigma$, 
and thus the scale of right handed neutrino masses, is 
smaller than the other VEVs. 
E.g., this happens when $x\sim 1/3$, that is $3 \lambda M \sim -2 \eta m$.

%% There is number of interesting cases: (1)~If $x\sim 0$ ($\lambda M\sim 3
%% \eta m$), the breaking is in two steps. In the first step, SO(10)$\to$L-R,
%% and right-handed neutrinos are massless.  (2)~If $x\sim 1/3$ ($3 \lambda M
%% \sim -2 \eta m$), there is a single step breaking due to $p\sim -a \sim
%% \omega =-m/3 \lambda$ but $\sigma$ remains small. (3)~The most
%% typical situation however is the one of single step breaking. E.g., fixing
%% $\lambda M=\eta m$, we get $p\sim-0.27\; m/\lambda$, $a\sim-0.67\;
%% m/\lambda$, $\omega \sim -0.21\; m/\lambda$, $\sigma \sim 0.51
%% m/\sqrt{\eta \lambda}$.  This corresponds to an approximate single step
%% breaking of SO(10)$\to$MSSM (but again if $\eta\gg \lambda$, the scale of
%% right-handed neutrinos can be somewhat lowered).

%% Vanishing of the D-terms (which implies $\sigma=\bar\sigma$) and the 
%% F-terms determines one of the degenerate (as usual in supersymmetric GUTs) 
%% vacua to be the broken left-right (L-R) symmetry chain 
%% \bea
%% p= a = \frac{1}{\sqrt{2}} \omega = -\frac{M}{10\eta} \, ; \quad \sigma^2 =
%%  \frac{P}{\eta^2}\left( \frac{3}{5}\lambda M - 2 \eta m \right)
%% \eea
%% giving an  approximate single-step breaking from SO(10) down to MSSM. 
%% Of course, it is possible to imagine intermediate mass scales with P-S 
%% and L-R symmetries; their detailed study is left for a forthcoming 
%% publication. 

Before proceeding with the inclusion of the light Higgs and matter 
superfields, we discuss the situation  regarding R-parity. 
Since it is equivalent to matter parity, 
$M \equiv (-1)^{3(B-L)}$, we discuss $M$ in what follows. Under $M$, 
${\bf 16}$ is odd and the rest even. 
On the other hand, SO(10) has a $Z_4$ center: 
${\bf 16}\to i{\bf 16}$,
${\bf 10}\to -{\bf 10}$, 
${\bf  210} \to {\bf 210}$, 
${\bf 126} \to -{\bf 126}$, 
$\overline{{\bf 126}} \to -\overline{{\bf 126}}$. 
Clearly, $M\in Z_4$ and thus R-parity is an automatic consequence of
gauge SO(10). Since $\Phi, \Sigma$ and $\overline\Sigma$ are even under 
$M$, R-parity continues to be an exact symmetry so far.

The ``light'' Higgs field we choose to be ${\bf 10}$-dimensional ($H$), and 
the superpotential gets an additional piece
\beq
\Delta W_H = m_H H^2 + \frac{1}{4!} \Phi_{ijkl} H_m (\alpha
\Sigma_{ijklm} + \beta \overline{\Sigma}_{ijklm})
\eeq
with three more complex parameters, seven in total.

With the matter fields being 16-dimensional spinors $\Psi_a$ ($a$=1,2,3
is a generation index), we have two symmetric Yukawa couplings in 
generation space 
\beq
W_Y = \Psi (Y_{10} H + Y_{126} \overline\Sigma) \Psi\;.
\eeq
We diagonalize one of them, 
which amounts to 3 real couplings; the other one then has $6\times 2=12$ 
more real parameters. 

With four Higgs superfields we can redefine four phases of the 
seven complex couplings and so we have $7\times 2-4=10$ real parameters 
in the Higgs superpotential $W_H$. Together with the Yukawa and gauge 
sector, this means $15+10+1 $, that is:
\begin{center}
{\em 26   real parameters in minimal {\rm SO(10)},}
\end{center}
just as in the MSSM. This is a remarkable result. Furthermore, 
the theory is in perfect accord with all the quark and lepton 
masses and mixings, as we now discuss.

{\it C. Fermion masses and mixings}\hspace{0.5cm} 
Naively, one imagines only the 
field $(2,2,1)$ in $H$ to have a VEV at the electroweak scale. 
%This is wrong for two reasons: it predicts vanishing 
%quark and lepton mixings, and equal masses of all charged leptons and 
%down quarks at the GUT scale. The former problem can be fixed by 
%introducing more {\bf 10}'s (but this is not a minimal theory anymore), 
%however the latter problem (the failure for the first two generations) 
%persists. 
The situation is more subtle. It is easy to see that 
$(2,2,15)$ fields in $\overline{\bf 126}$ mix with 
$(2,2,1)$ through the VEV of $(1,1,15)$ in ${\bf 210}$, which is 
of order $M_{GUT}$. The usual minimal fine tuning implies that 
one combination of the two 
L-R bidoublets is light and thus 
both $(2,2,1)$ and $(2,2,15)$ contribute to fermion masses. 
The idea to use ${\bf 210}$ for this purpose was suggested long time 
ago \cite{Babu:1992ia} 
and the predictions for the fermions masses were 
studied extensively 
\cite{Babu:1992ia,Oda:1998na,Brahmachari:1997cq}. It was never 
realized though that  ${\bf 210}$ alone is sufficient. What 
seemed to be an ugly model building, where one chooses the fields and 
interactions for a particular scope, turns out to be the minimal 
supersymmetric SO(10), or, better to say, the minimal supersymmetric 
grand unified theory. 

What is really surprising is that the theory is still consistent with all 
the data even when one specifies the nature of the see-saw mechanism 
\cite{gmrs,yana,Mohapatra:1979ia,glashow,wet,gm}. As is well known, 
the see-saw may proceed through the right-handed 
neutrino masses (type I see-saw) or through the VEV of the 
$(3,1,\overline{10})$, which necessarily gets induced (type II see-saw). 
The two limiting cases mean one parameter less and yet both work and both 
predict the 1-3 leptonic mixing angle quite large: $0.16$ 
\cite{Matsuda:2001bg,Goh:2003sy}. The type II case is particularly 
interesting, since the large atmospheric mixing angle is intimately 
tied to $b-\tau$ unification \cite{Bajc:2002iw}. 

{\it D. The fate of R-parity}\hspace{0.5cm} 
The theory is even more predictive. It uniquely determines the 
low-energy effective theory: it is MSSM with exact R-parity. 
As we have seen, R-parity remains unbroken at the first stage of symmetry 
breaking. The question is, what happens at the low energy supersymmetry breaking or 
electroweak scale. More precisely, one must know whether light sneutrinos 
get a VEV and thus break R-parity. This seems a hopeless task since it 
should depend on the theory of supersymmetry breaking and soft masses. 
Fortunately, a spontaneous breakdown 
of R-parity through the sneutrino VEV would result 
in the existence of a pseudo-Majoron with its mass inversely proportional 
to the right-handed neutrino mass. This is ruled out by the Z decay width 
\cite{Aulakh:1997ba,Aulakh:1999cd}. This is completely analogous to the 
impossibility 
of breaking R-parity spontaneously in the MSSM, where the Majoron is 
strictly massless. 

{\it E. Proton decay}\hspace{0.5cm} 
Another important aspect of the theory regards proton decay. 
For simplicity, here 
we assume a single step breaking from the GUT 
symmetry down to the MSSM. The GUT scale then is, neglecting the GUT 
threshold effects, close to $10^{16}$ GeV 
\cite{Dimopoulos:1981yj,Ibanez:yh,Einhorn:1981sx,Marciano:1981un}. 
In this case $d=6$ induced proton decay is on the 
slow side and will be a long time before being verified 
(for a recent attempt to increase this rate see for example 
\cite{Klebanov:2003my}). 
On the other hand the $d=5$ decay tends to be quite fast 
\cite{Sakai:1981pk,Weinberg:1981wj}. For example in minimal 
SU(5) the situation is dramatic 
\cite{Hisano:1992jj,Goto:1998qg,Murayama:2001ur}, 
although there are ways out 
\cite{Bajc:2002bv,Bajc:2002pg,Emmanuel-Costa:2003pu}. 
This will clearly be one of the 
central tests of the theory. What are the ingredients that go into the 
analysis? 
First of all, there are a number of colour triplet supermultiplets 
responsible for the decay, as in other SO(10) model \cite{Babu:1997js}. 
The effective rate depends on their masses and mixings, and also the 
mixings to other colour triplets, which do not directly couple to 
fermions. Those colour triplets that contribute directly live in 
${\bf 10}$ and ${\bf \overline{126}}$, whereas the others that just 
mix with them live in ${\bf 210}$ and ${\bf 126}$. The detailed 
discussion is beyond the scope of this note, but it is extremely 
important and is being performed in  detail. A related issue 
is the existence of intermediate symmetry breaking scales which in 
principle affect the triplet masses. 

{\it F. Leptogenesis}\hspace{0.5cm} 
An important appealing feature of the see-saw mechanism in general 
and SO(10) in particular is the celebrated mechanism of leptogenesis 
\cite{Fukugita:1986hr}. It would be nice to see how it works in this 
theory. The situation is quite complicated though. In most works 
one assumes the type I see-saw mechanism being responsible both for 
neutrino masses and leptogenesis. This of course does not have to be true 
at all and we have a number of possibilities depending of what dominates 
the neutrino masses and what is responsible for leptogenesis. Although
the constraints from leptogenesis can not be taken as seriously 
as those from proton decay and even more fermion masses, 
certainly they are worth studying.

{\it G. Conclusions}\hspace{0.5cm} 
In this letter we have presented what is according to us the minimal 
supersymmetric grand unified theory. In defining minimality we have used 
the criterion of the economy of the parameters or, equivalently, the power 
of predictability. We have sticked to the gauge principle, and demanded 
the 
simple gauge group to be the only symmetry of the theory. This enables one 
to have a well defined framework, which can test the idea of 
supersymmetric grand unification. In a sense, this minimal theory should be 
viewed as the GUT analogue of the standard model as the minimal theory of 
low-energy electro-weak phenomena. 

The theory is based on the SO(10) gauge symmetry and contains the usual 
three generations of ${\bf 16}$-dimensional matter superfields, a (light) 
${\bf 10}$-dimensional Higgs superfield and the (heavy) ${\bf 210}$, 
${\bf 126}$ and ${\bf\overline{126}}$ representation Higgses. The theory 
is blessed by a small number of parameters, 26 real ones in total, which 
makes it quite predictable. One often fears large representations, 
according to us for no valid reason. In fact large representations often 
mean less parameters. They also have an important characteristic of 
predicting the Landau pole above the physical scale in question. In this 
particular case this happens at the scale $\Lambda_F$ and order of 
magnitude or so above $M_{GUT}$. Someone could be unhappy about this, but 
to us this is an interesting prediction of the theory: there is new, 
presumably strongly coupled dynamics not far from the unification scale. 
The scale $\Lambda_F$ as much as $M_{Pl}$ could easily leave imprints 
at the GUT scale. We decided to ignore them in order to be concrete. 
The theory makes clear predictions and experiment will tell us how good 
they are. One could also imagine a scenario with $\Lambda_F\sim
M_{GUT}$, in which case the theory becomes strongly coupled \cite{c}.

Contrast this with what is often coined the minimal supersymmetric 
SO(10) theory based on ${\bf 16}_H$, $\overline{\bf{16}}_H$, 
${\bf 45}_H$, ${\bf 10}$ and non-renormalizable superpotentials 
\cite{Babu:1994dc,Babu:1994kb,Dvali:1996wh,Barr:1997hq,Chacko:1998jz}. 
At the order $1/M_{Pl}$ this theory has many more parameters; for example, 
it has 4 sets of Yukawa couplings (33 real parameters in total). This is 
clearly much less economical than the theory we presented. Obviously 
choosing particular textures is against the spirit of extracting 
information from the principle of grand unification; it has to do much more 
with flavour physics, which is outside grand unification. On top of that, 
it has to be augmented with extra discrete (or continuous) symmetries 
in order not to break R-parity at $M_{GUT}$. 

We have shown that there is consistent single-step symmetry breaking 
down to the MSSM. The main outcome is the exact R-parity which guarantees 
the stability of the lightest supersymmetric partner and its possible role 
as the dark matter. Furthermore, the theory is completely in accord with 
all the data on fermion masses and mixings. In the limiting cases of type 
I or type II see-saw it predicts the 1-3 leptonic mixing angle to be about 
$0.16$; whether this remains true for the general see-saw formula remains 
still to be seen. If it survives, it could be the smoking gun of the 
model. As we discussed above, another important test of the theory will be 
provided by the careful evaluation of proton decay, baryon asymmetry of 
the universe and flavour violating processes. We plan to address these 
questions in a future publication. 

\acknowledgments

The work of A.M., G.S. and B.B.\ is supported by CDCHT-ULA 
(Project C-1073-01-05-A), EEC (TMR contracts ERBFMRX-CT960090 and 
HPRN-CT-2000-00152) and the Ministry of Education, Science 
and Sport of the Republic of Slovenia, respectively. A.M. and F.V.\ thank 
ICTP, and INFN exchange program (F.V.), for hospitality during the course 
of this work. We thank Z.~Berezhiani, G.~Dvali and
R.N.~Mohapatra for useful discussions.

\end{document}